# Breakdown of Interference Rules in Azulene, a Non-Alternant Hydrocarbon

Jianlong Xia,[1,‡] Brian Capozzi,[2,‡] Sujun Wei,[1] Mikkel Strange,[3] Arunabh Batra,[2] Jose R. Moreno,[1] Roey J. Amir,[4] Elizabeth Amir,[5] Gemma C. Solomon,[3] Latha Venkataraman,*[,2] Luis M. Campos.*[,1]

[1] *Department of Chemistry, and* [2]*Department of Applied Physics and Applied Mathematics, Columbia University, New York, NY 10027, USA,*
[3] *Nano-Science Center and Department of Chemistry, University of Copenhagen, Copenhagen 2100, Denmark*
[4] *School of Chemistry, Tel-Aviv University, Tel-Aviv 69978, Israel.*
[5] *The Pernick Faculty of Engineering, Shenkar College of Engineering and Design, Ramat-Gan 52526, Israel.*

‡ These authors contributed equally.

* Corresponding Author Email: lv2117@columbia.edu, lcampos@columbia.edu

ABSTRACT: We have designed and synthesized five azulene derivatives containing gold-binding groups at different points of connectivity within the azulene core to probe the effects of quantum interference through single-molecule conductance measurements. We compare conducting paths through the 5-membered ring, 7-membered ring, and across the long axis of azulene. We find that changing the points of connectivity in the azulene impacts the optical properties (as determined from UV-Vis absorption spectra) and the conductivity. Importantly, we show here that simple models cannot be used to predict quantum interference characteristics of non-alternant hydrocarbons. As an exemplary case, we show that azulene derivatives that are predicted to exhibit destructive interference based on widely accepted atom-counting models show a significant conductance at low biases. Although simple models to predict the low-bias conductance do not hold with all azulene derivatives, we show that the measured conductance trend for all molecules studied actually agrees with predictions based on the more complete GW calculations for model systems.

Due to the vast potential applications of organic semiconductors, molecules with pi-conjugated systems have been studied extensively to understand their charge transport and optical properties. In fact, alternant hydrocarbons can be readily synthesized and their physicochemical properties can be finely tuned. Through rational chemical design, highly conductive molecular wires have been realized,[1] and recent scientific developments have made possible the fabrication of molecular analogues of several components, such as transistors,[2-4] potentiometers,[5] diodes[6-9] and switches.[10-13] Such modularity arises from the strong correlation between conductance, molecular structure, and connectivity.[4, 14, 15] While conductance and interference patterns in alternant hydrocarbons have been the main focus,[14-19] little is known about how substitution patterns affect the single-molecule conductance properties of non-alternant hydrocarbons,[20] where quantum interference (QI) effects may not be predicted based on simple bond-counting models.[21, 22]

The optical properties and chemistry of alternant and non-alternant hydrocarbons vary widely with respect to each other.[20] A classical example is the comparison of naphthalene to azulene. Azulene has several unique characteristics that distinguish it from its aromatic bicyclic $C_{10}H_8$ isomer, naphthalene. In its simplest form, resonance delocalization of azulene yields an electron-deficient 7-membered ring, and an electron-rich 5-membered ring to achieve Hückel aromatic stabilization – the source of its inherent dipole moment. Both naphthalene and azulene exhibit dramatically different optical properties; for example, naphthalene is white and azulene is a blue compound.[23] The optical transition in azulene that renders it blue in color occurs from the energy gap between the $S_1$-$S_2$ transition in the visible spectrum,[24, 25] as opposed to the $S_0$-$S_1$ transition, which is the main transition of naphthalene in the UV region (ca. 280 nm). Such phenomena can be explained by a molecular orbital (MO) picture.[20, 23] The highest occupied MO (HOMO) of naphthalene and the lowest unoccupied MO (LUMO) occupy the same space, whereas these orbitals have different nodes in azulene, resulting in little overlap and a net electron repulsive energy (Figure 1A). Considering the MOs, substitution patterns of electron donating/withdrawing groups lead to the predictive power and tunability of optical transitions by altering the stability of the MOs at specific sites and in particular patterns.[25] Consequently, the single-molecule conductance patterns through naphthalene can also be reasoned with the MO picture and tight-binding energy transmission curves.[14, 22] However, little is known about the effects of the substitution patterns of azulene on the single molecule conductance by "wiring" azulene through various points of connectivity (Figure 1B).

Here, we use an experimental approach involving modular synthesis of azulenes and scanning tunneling microscope-break junction (STM-BJ) measurements, along with a theoretical approach to demonstrate the relation between interference effects and conductance through various points of pi-bond conjugation in azulene, a prototypical non-alternant hydrocarbon. We couple gold-binding groups based on thiochroman[26] to azulene at the 1,3-, 2,6-, 4,7-, and 5,7- positions, yielding four molecules with different conjugation patterns that couple directly to carbons having different orbital coefficients in both the HOMO and LUMO (Figure 1). In addition, we synthesize a control pyridine functionalized azulene with connections at the 1,3 position (1,3PyAz) to support our studies. Our recent advancements to make functional azulenes[27] made possible the synthesis of 4,7Az, 5,7Az and 2,6Az in only a few steps, whereas the synthesis of 1,3Az was accomplished by standard aromatic coupling strategies, from 2,3-dibromoazulene[28] as detailed in the Supporting Information (SI) document.

We use the STM-BJ technique under ambient conditions[29] to measure the electronic transport characteristics of these compounds in solution. We compared the conductance of single-molecule junctions created with these azulene derivatives that connect to Au electrodes through one of the two rings, or across the molecular backbone. Additionally, the conductance trends in naphthalene were used as a model system. naphthalene obeys fundamental rules: the 2,7Np and 1,3Np derivatives do not conduct while the 2,6Np and 1,4Np show clear conductance signatures.[14, 22] These observations for naphthalene are consistent with the atom-counting model by Markussen et al.,[21] which predicts QI for the 2,7Np and 1,3Np derivatives. Interestingly, the same model would predict that 1,3Az, and 5,7Az should exhibit QI (see Figure 1). Our experiments and calculations reported here are aimed at determining if QI rules hold with non-alternant hydrocarbons, using azulene as a model system.

In Figure 2, we compare the UV-Vis spectra for the four thiochroman-linked compounds synthesized. The $S_0$-$S_1$ transition between the lowest energy vibrational states (*ie.* 0-0 transition bands) for 5,7Az, 1,3Az, and 4,7Az are at ca. 750 nm (1.65 eV), 720 nm (1.72 eV), and 720 nm (1.72 eV), respectively, whereas for 2,6Az, it is slightly higher in energy at 700 nm (1.77 eV). These short shifts observed are in agreement with the expected substitution pattern of electron donating substituents at their respective positions,[25] and the trends can be useful since they scale with the fundamental energy gap (i.e. the difference between the ionization potential electron affinity).[30] The marked differences in optical changes are more evident from the $S_0$-$S_2$ transition: 5,7Az has the most blue-shifted optical onsets of wavelength absorption ($\lambda_{onset}$) at 390 nm (3.18 eV), and both 1,3Az and 4,7Az have similar $\lambda_{onset}$ values of 410 nm and 420 nm, respectively, corresponding to an energy gap of approximately 3.0 eV, which is lower in energy by 0.5 eV than that of the singlet $S_0$-$S_2$ transition in azulene.[31] This trend has been observed in other conjugated azulenes with similar substitution patterns arising from destabilization of the HOMO and LUMO+1.[27, 28] Notably, 2,6Az has a much lower $S_0$-$S_2$ transition energy onset ca. 460 nm (2.70 eV), arising from conjugation of the electron-donating gold-binding groups through the long axis of the dipole.[32] In other aromatic conjugated oligomers containing a single azulene unit, the $\lambda_{onset}$ does not red-shift as much as it does in the case of 2,6Az. As it will be discussed in the following section, these expected trends in optical changes do not translate to predictable trends in single-molecule conductance of this non-alternant hydrocarbon, unlike measurements of other systems such as naphthalene.

To measure molecular conductance, single-molecule junctions are created by repeatedly forming and breaking Au point contacts with a modified STM in a 1,2,4-trichlorobenzene solution of the molecules.[29] Thousands of conductance versus displacement traces are measured with each molecule (Figure 3a). The traces reveal steps at molecule-dependent conductance values below $G_0$, the quantum of conductance, that are due to conduction through a molecule bonded in the gap between the two Au point-contacts. Repeated measurements give a statistical assessment of the junction properties. Figure 3b shows one-dimensional conductance-histograms generated from thousands of traces without data selection using logarithm bins.[33] The most frequently measured conductance value of each molecule is given by the peak value of these histograms.

The histograms show a clear conductance peak for all four thiochroman-linked azulene derivatives, and the magnitude of the conductance depends on the points of connectivity. To confirm that these measurements do represent junctions formed by trapping the azulenes at the linkers, we also create two-dimensional conductance-displacement histograms (see SI Figure S1). These preserve the distance that single-molecule junctions can be elongated, which is proportional to the length of the molecule being sampled; this provides a secondary probe of junction formation.[34] We use the 2D histograms to determine an average conductance plateau length for each molecule studied and tabulate these results along with conductance values (from Figure 3b) in Table 1. We see that as the molecular backbone length increases, the conductance plateau length also increases indicating that we are indeed measuring transport through these molecules. However, we note here that the conductance values do not correlate with the molecular energy gap determined from the UV-Vis spectra (Figure 2b). More importantly, these conductance histograms show a clear peak for the two molecules that are predicted to display destructive interference effects, namely the 1,3Az and 5,7Az. One possible explanation for this result is that both the 1,3Az and 5,7Az are conducting through the sigma system, however, this would not explain the order of magnitude difference between the conductance of 1,3Az and 5,7Az, as conductance is occurring through the same number of bonds. Also, conductance through the sigma systems of these two molecules we be predicted to give significantly lower conductance values.[26] Furthermore, we note that the 5,7- derivative displays a secondary conductance feature at ~$10^{-3}$ $G_0$ in the histogram in Figure 3b. Based upon the short elongation length of this feature (Figure S1), we conclude that it does not correspond to a molecule bound across the junction with both terminal sulfurs. Instead, we hypothesize that it relates to a junction bound on one side by the linker and on the other side by an interaction of the azulene pi-system with the Au electrode.[5]

To determine if the strong conductance signature measured in the 1,3Az is a robust result, we synthesized and measured a pyridine-terminated derivative (1,3PyAz). The 1,3PyAz shows a clear double conductance peak, characteristic of all other pyridine-linked molecules measured indicating a coupling through the azulene pi-system. The conductance of 1,3PyAz is lower than that of the thiochroman analog ($G_0$ = 9 x $10^{-5}$, table 1, Figure S2), however this is consistent with previous observation. Specifically, the conductance of 4,4'-bipyridine is lower than that of a biphenyl with amine or thiochroman linkers,[13] and these differences in conductance are explained by considering the orbital energy level alterations induced by the pyridyl linkers, an effect that is also evident in the expected absorption shift in the UV-vis (Figure S3). Given that we observe conductance in both 1,3Az and 1,3PyAz, we reasoned that simple atom counting models that explain conductance do not hold with non-alternant hydrocarbons. Therefore, to better understand why the conductance does not correlate with the optical energy gaps, and to see how the effects of interference play a role in the transport of these systems, we turn to calculations.

The charge transport characteristics of azulene and naphthalene are compared using a model molecular system with the azulene and naphthalene backbone connected to ethynyl linkers. These simplified models were used here as we expect that only the azulene core (and the attachment sites for the linkers) will dictate whether or not destructive interference effects dominate transport through the molecule. Furthermore, the GW model Hamiltonian treats only the pi-systems of the molecules considered, and thus changes to the linker structures should not affect the trends determined in these calculations. We minimize the geometry for these model systems using density functional theory. We then describe exchange and correlation effects in the GW approximation,[35, 36] attach the molecules to featureless, wideband leads and calculate the transmission using the Landauer-Büttiker formula (see SI for more details).[37] The transmission curves for various substitution patterns in naphthalene and azulene are shown in Figure 4. We see first that the HOMO and LUMO transmission resonances located near -3eV and 3eV in naphthalene are symmetric about the Fermi Energy ($E_F$) in contrast to azulene. Symmetric curves are common for alternant hydrocarbons where HOMOs and LUMOs are paired, as opposed to non-alternant hydrocarbons, which have unpaired frontier orbitals.[38] The charge gap (HOMO-LUMO gap) is seen to change slightly with substitution pattern, but is in general larger for naphthalene than for azulene consistent with the UV-Vis results.

We now compare, in detail the transmission spectra for the naphthalene and azulene derivatives noted in Figure 4. For 1,3Np and 2,7Np, we see that transmission is strongly suppressed in the middle of the gap and shows an $E^2$ energy dependence, characteristic of destructive QI[19, 39] in contrast to the other naphthalene derivatives which do not show any destructive QI within the HOMO-LUMO gap. These results for naphthalene agree with the simple interference atom counting rule, which predicts a transmission zero at $E = 0$ eV for the 1,3Np and 2,7Np derivatives as shown in Figure 1b. An important difference in the transmission functions of the azulene derivatives is that they are non-zero at $E = 0$ V. Instead, the plot shows two transmission zeros for 1,3Az and 5,7Az within the HOMO-LUMO gap, located near the LUMO transmission resonance. This means that even though the transmission for 1,3Az and 5,7Az show destructive interference features, these do not occur at $E_F$. The resulting transmission at $E \sim 0$ eV is quite large for 1,3Az and, in fact, is even higher than all other derivatives for $E < -1$ eV. We note that the basic tight-binding model, from which the atom counting method was derived, also shows a non-zero transmission value at $E = 0$ eV (see SI).

Since these transmission plots are determined from model junctions without including the electron-donating/withdrawing effects of the linkers, we cannot determine where $E_F$ is located in the real systems studied in the experiment. We can however infer that $E_F < 0$ for the thiochroman linkers, while $E_F > 0$ for the pyridine linker.[13] Considering $E_F = -1.5$ eV, the conductance trends of the azulene molecules follow the order: 1,3Az > 2,6Az ~ 4,7Az > 5,7Az. Such a trend is in qualitative agreement with the experimental results. For the 1,3PyAz, we would consider $E_F$ above 0, which would still agree with the trend that the conductance of 1,3Az is greater than that of 1,3PyAz. We stress that only the relative conductance trend should be compared with the experiments, since the models for the calculations were simplified by excluding the large thiochroman linkers, among other simplifications (see SI).

In summary, we have analyzed the various paths of conduction through different points of connectivity in five azulene derivatives. Our key finding illustrates that simple atom-counting models do not describe interference effects correctly around the metal Fermi energy due to the asymmetric eigen-spectrum in these non-alternant hydrocarbons. We show however, through GW based transmission calculations, that we can rationalize our findings and the conductance trends for both HOMO- and LUMO-conducting molecules. Understanding the fundamental conductance characteristics of azulenes provide insights of this molecular building block that can be used to synthesize optoelectronic materials.[40] Collectively, single-molecule and bulk material studies can contribute to our understanding of the physical properties across different length scales to design materials for next-generation devices. Furthermore, our results could be built on to formulate a general understanding of conductance through non-alternant hydrocarbon systems by carrying out single-molecule conductance measurements on other hallmark non-alternant hydrocarbons such as fulvene and cyclopropene.

**Acknowledgements.** This work was supported by the National Science Foundation, under grant number DMR-1206202. GCS and MS received funding from the European Research Council under the European Union's Seventh Framework Programme (FP7/2007-2013)/ ERC Grant agreement no. 258806.

**Supporting Information Available**. Synthetic procedures, single-molecule measurement details, and computational information can be found free of charge via the Internet at http://pubs.acs.org.

**Notes.** The authors declare no competing financial interest.

# Table 1

| Molecule | Calculated Length (nm) | 2D Histogram Step Length (nm) | Average Conductance ($G_0$) |
|---|---|---|---|
| 1,3Az | 1.36 | 0.61 | $32 \times 10^{-5}$ |
| 1,3PyAz | 1.03 | 0.56 | $9 \times 10^{-5}$ |
| 2,6Az | 1.74 | 0.79 | $32 \times 10^{-5}$ |
| 4,7Az | 1.50 | 0.71 | $8 \times 10^{-5}$ |
| 5,7Az | 1.31 | 0.70 | $2 \times 10^{-5}$ |

**Table 1.** Single-molecule conductance values of the four thiochroman-linked azulene derivatives and the pyridine-linked azulene (1,3PyAz), along with the calculated end-to-end conductance

path length determined from DFT optimized structures using (B3LYP/6-31G*), and the measured step-length from the 2D histograms shown in Figure S1 (Supporting Information). Details on the measured step-length can be found in the caption of Figure S1.

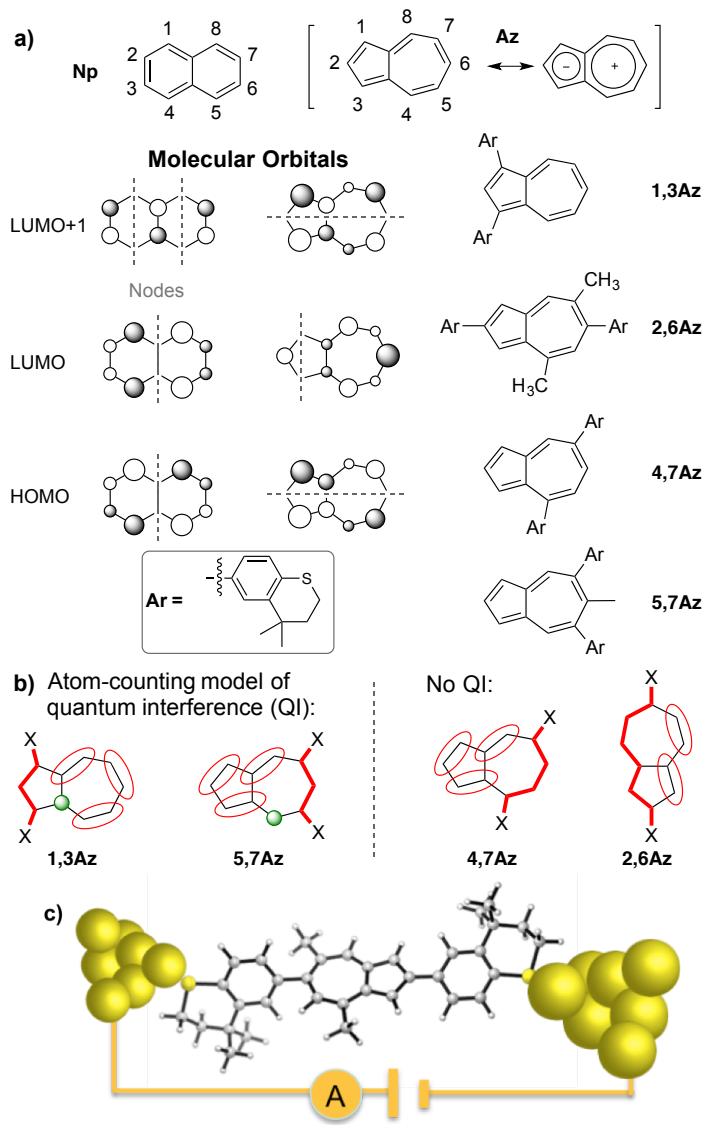

**Figure 1. a)** Molecular orbitals of azulene and naphthalene, along with the four derivatives of interest for single-molecule conductance studies. **b)** Atom-counting model that can be used to predict quantum interference for molecules where the conducting path is identified and the remaining atoms of the pi-system are not paired, as shown in the picture (green dots). Similarly, no interference is predicted when the remaining atoms are all paired after identifying the conductance path (in bold). **c)** Schematic representation of the azulene derivative 2,6Az in a scanning tunneling microscope-break junction.

Figure 2

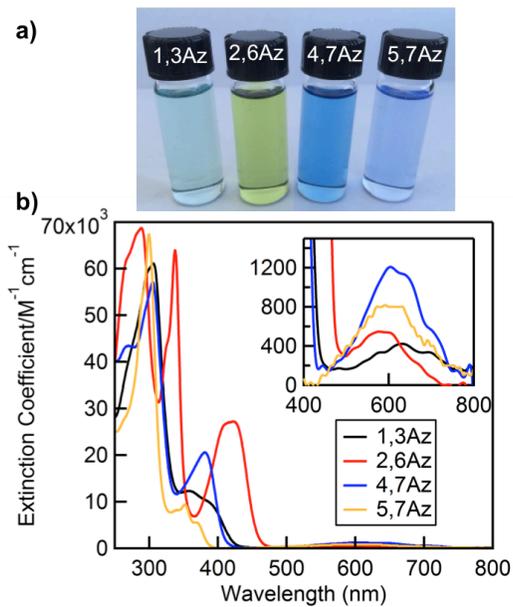

**Figure 2. a)** Photograph of 1,3Az, 2,6Az, 5,7Az and 4,7Az in dichloromethane. **b)** UV-vis spectra of the four samples. Inset: magnification of the region between 400 and 800 nm ($S_0$-$S_1$ transition). All measurements were carried out on solutions of 10μM concentration.

Figure 3

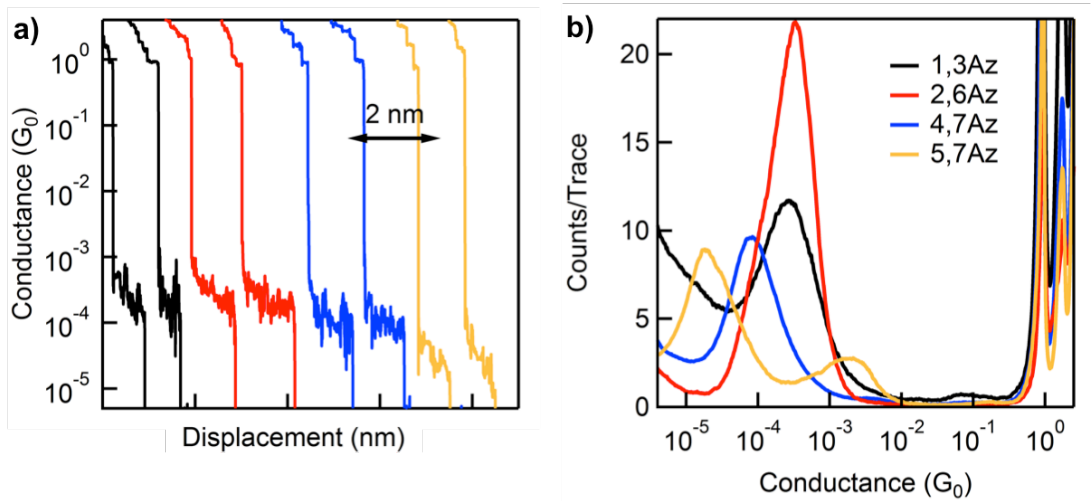

**Figure 3. a)** Samples of single conductance-displacement traces offset laterally for clarity. **b)** Log-binned 1D conductance histograms using 100 bins/decade that are measured at 250 mV. The histograms for 1,3Az, 4,7Az, and 5,7Az are each created from 10,000 individual conductance traces, while the histogram for 2,6Az is created from 5,000 conductance traces.

# Figure 4

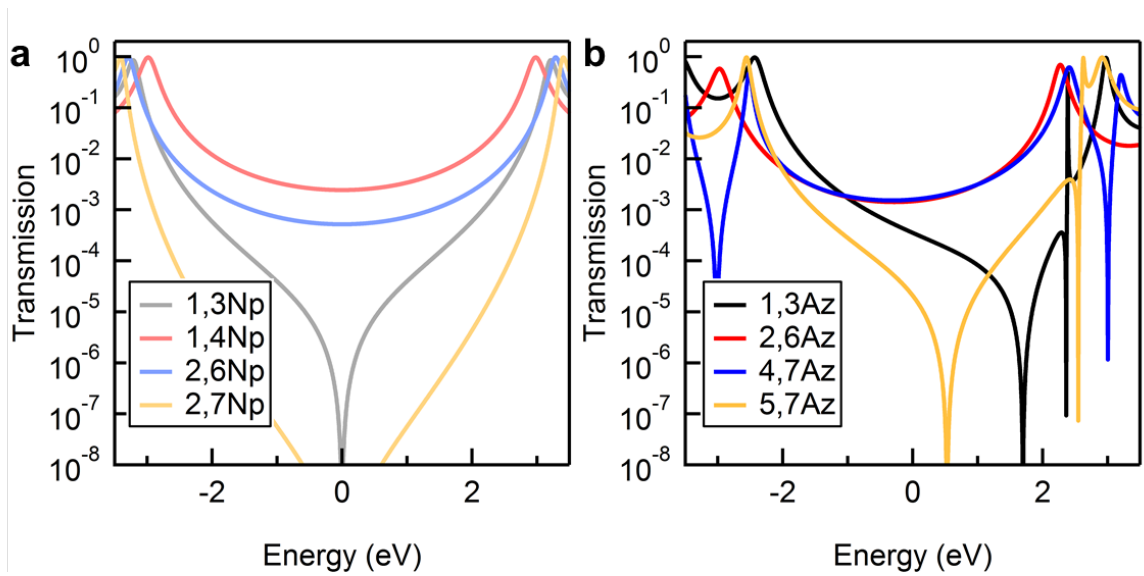

**Figure 4.** Transmission for various linker substitution patterns for **a)** Naphthalene and **b)** azulene obtained from GW calculations.